\documentclass[cits]{PoS}\usepackage{amsmath} 
\title{Mass of the $b$ Quark from QCD Sum Rules for $f_{B_{(s)}}$}
\ShortTitle{Mass of the $b$ Quark from QCD Sum Rules}
\author{Wolfgang Lucha\\Institute for High Energy Physics,
Austrian Academy of Sciences, Nikolsdorfergasse 18, A-1050 Vienna,
Austria\\E-mail: \email{Wolfgang.Lucha@oeaw.ac.at}}
\author{\speaker{Dmitri Melikhov}\\Institute for High Energy
Physics, Austrian Academy of Sciences, Nikolsdorfergasse 18,
A-1050 Vienna, Austria,\\Faculty of Physics, University of Vienna,
Boltzmanngasse 5, A-1090 Vienna, Austria, and\\D.~V.~Skobeltsyn
Institute of Nuclear Physics, Moscow State University, 119991,
Moscow, Russia\\E-mail: \email{dmitri\_melikhov@gmx.de}}
\author{Silvano Simula\\INFN, Sezione di Roma Tre, Via della Vasca
Navale 84, I-00146 Roma, Italy\\E-mail:
\email{simula@roma3.infn.it}}

\abstract{We demonstrate that Borel QCD sum rules for heavy--light
currents entail a very strong correlation between the $b$-quark
mass $m_b$ and the $B$-meson's decay constant $f_B,$ that is,
$\delta f_B/f_B\approx-8\,\delta m_b/m_b.$ By starting from $f_B$
as input, this observation allows for an accurate sum-rule
determination of $m_b.$ Employing precise lattice QCD results for
$f_B$ in our sum-rule study based on the three-loop $O(\alpha_{\rm
s}^2)$ heavy--light correlation function implies
$\overline{m}_b(\overline{m}_b)=(4.247\pm0.034)\;\mbox{GeV}$ for
the $b$-quark~$\overline{\rm MS}$ mass.}

\FullConference{The European Physical Society Conference on High
Energy Physics -- EPS-HEP2013\\18--24 July 2013\\Stockholm,
Sweden}

\begin{document}\section{Introduction}Within the ``standard model
of elementary particle physics,'' the mass of the $b$ quark
constitutes a fundamental parameter of the theory. Therefore, the
knowledge of its numerical value as precisely as possible is of
utmost importance. Lattice QCD provides a framework to determine
this parameter by direct albeit purely numerical procedures;
unfortunately, the $b$ quark is too heavy to be dealt with by
current lattice setups: lattice-QCD computations of its mass
require either an extrapolation of the lattice-QCD findings from
lighter simulated masses or the use of the ``heavy-quark effective
theory'' (HQET) formulated on the lattice. The actual value of a
quark mass depends on the renormalization scheme employed for the
rigorous definition of this quantity; for the $b$ quark, usually
the predictions for its pole mass, for its $\overline{\rm MS}$
running mass at renormalization scale $\nu,$
$\overline{m}_b(\nu),$ or for
$m_b\equiv\overline{m}_b(\overline{m}_b)$ are compared. Using
unquenched gauge configurations and $N_f=2$ dynamical sea-quark
flavours~gives:\begin{itemize}\item
$m_b=(4.29\pm0.14)\;\mbox{GeV}$ \cite{ETMC1} and
$m_b=(4.35\pm0.12)\;\mbox{GeV}$ \cite{ETMC2} when confiding in
extrapolation;\item $m_b=(4.26\pm0.09)\;\mbox{GeV}$
\cite{mb_Gimenez}, $m_b=(4.25\pm0.11)\;\mbox{GeV}$ \cite{mb_UKQCD}
and $m_b=(4.22\pm0.11)\;\mbox{GeV}$ \cite{ALPHA}, for instance, if
one is willing to accept the expansions involved in the HQET-based
computations.\end{itemize}Moment sum rules for two-point functions
of heavy--heavy currents entail more accurate $m_b$ values:
\begin{itemize}\item Low-$n$ moment sum rules adopting
three-loop $O(\alpha_{\rm s}^2)$ \cite{mb4} and four-loop
$O(\alpha_{\rm s}^3)$ \cite{mb1} fixed-order perturbative-QCD
results combined with experiment yield
$m_b=(4.209\pm0.050)\;\mbox{GeV}$ \cite{mb4} and
$m_b=(4.163\pm0.016)\;\mbox{GeV}$ \cite{mb1}, respectively, where
the latter result is supported by combining perturbative QCD and
lattice-QCD efforts using $N_f=2+1$ dynamical sea-quark~flavours
\cite{mb2}.\item Large-$n$ moment renormalization-group-improved
next-to-next-to-leading-logarithmic-order $\Upsilon$ sum rules,
underpinned by experiment, give
$m_b=(4.235\pm0.055_{\rm(pert)}\pm0.03_{\rm(exp)})\;\mbox{GeV}$
\cite{hoang}.\end{itemize}Our recent study of $m_b$ by means of
heavy--light QCD sum rules reveals that $m_b$ may be found~with
comparable accuracy if a precise input value of the
$B_{(s)}$-meson decay constant $f_{B_{(s)}}$ is
available~\cite{lms2013}.

\section{Anticorrelation Between Beauty-Meson Decay Constant and
Bottom-Quark Mass}Quantum theory allows for easy exploration of
the sensitivity of any $B_{(s)}$-meson decay constant
$f_{B_{(s)}}$ to the $b$-quark mass $m_b$: in any nonrelativistic
potential model where the potential involves~only one coupling
constant (e.g., pure Coulomb or pure harmonic-oscillator
potentials), the ground-state wave function at the origin
$\psi(0)$ and binding energy $\varepsilon$ are related by
$|\psi(0)|\propto\varepsilon^{3/2};$ for any potential that is a
sum of confining and Coulomb interactions, this relation is only a
good approximation~\cite{lms_qcdvsqm}.

Recalling that a decay constant is the analogue of the wave
function at the origin and exploiting the (well-known) scaling
behaviour of the decay constant of a heavy meson in the
heavy-quark limit entails, as approximate relation between
$B$-meson mass $M_B$ and pole mass $m_Q$ of the heavy quark~$Q,$
$$f_B\,\sqrt{M_B}=\kappa\,(M_B-m_Q)^{3/2}\ .$$Now, keeping $M_B$
fixed and equal to its experimental value $M_B=5.27\;\mbox{GeV},$
we can easily derive the dependence of $f_B$ on small variations
$\delta m_Q$ around some given value of $m_Q.$ Taking into account
that $f_B\approx200\;\mbox{MeV}$ for
$m_Q\approx4.6\mbox{--}4.7\;\mbox{GeV},$ we obtain
$\kappa\approx0.9\mbox{--}1.0$ and $\delta f_B\approx-0.5\,\delta
m_Q$ or, equivalently,$$\frac{\delta
f_B}{f_B}\approx-(11\mbox{--}12)\,\frac{\delta m_Q}{m_Q}\ .$$From
this example, we expect a rather high sensitivity of $f_B$ to
$m_Q$: Varying $m_Q$ by $+100\;\mbox{MeV}$ gives $\delta
f_B\approx-50\;\mbox{MeV}.$ Similar effects should be observable
in the predictions of QCD sum rules~\cite{svz,aliev}.

\section{QCD Sum-Rule Extractions of Beauty-Meson Decay Constants
in the Literature}More or less recently, several QCD sum-rule
results \cite{nar2001,jamin,lms_fB,nar2012} of beauty-meson decay
constants using three-loop heavy--light correlators \cite{chet},
all of them deriving, in fact, from essentially the same
analytical expression for the correlator, have been published;
Table~\ref{Table:1} summarizes the corresponding findings for
$f_B.$ At first glance, the predictions appear to be rather stable
and practically independent of the $m_b$ input value. However,
this may not be put forward as argument in support of the
reliability of all the extractions, since, evidently, the figures
in Table \ref{Table:1} do not follow our above general~pattern.
For instance, the central values of $m_b$ found by
Ref.~\cite{nar2001} and Ref.~\cite{nar2012} differ by almost
$200\;\mbox{MeV}$~but the corresponding decay constants are nearly
identical. This forces us to suspect that not all findings are
equally trustable. Recall that the ground-state parameters in
Table \ref{Table:1} are subject to two decisions:\begin{itemize}
\item How is the three-loop perturbative result organized in terms
of pole or $\overline{\rm MS}$ heavy-quark mass?\item How are the
{\em auxiliary\/} sum-rule quantities, such as the {\em
effective\/} onset of the continuum, fixed?\end{itemize}

\begin{table}[h]\begin{center}\caption{QCD sum-rule extractions of
the $B$-meson decay constant $f_B$ from heavy--light two-point
correlator.}\label{Table:1}\vspace{2ex}\begin{tabular}{lcccc}
\hline\hline&Reference~\cite{nar2001}&Reference~\cite{jamin}&
Reference~\cite{lms_fB}&Reference~\cite{nar2012}\\\hline$m_b$
(GeV)&$4.05\pm0.06$&$4.21\pm0.05$&$4.245\pm0.025$&
$4.236\pm0.069$\\$f_B$ (MeV)&$203\pm23$&$210\pm19$&$193\pm15$&
$206\pm 7$\\\hline\hline\end{tabular}\end{center}\end{table}

In a recent critical detailed analysis of the sum-rule extraction
of $f_B$ \cite{lms2013}, we demonstrated~that,\begin{itemize}\item
if the correlator is expressed in terms of the $\overline{\rm MS}$
running instead of the pole $b$-quark mass~and\item if consistent
procedures for the extraction of the bound-state properties of
interest are applied,\end{itemize}the QCD sum-rule extractions of
$f_B$ exhibit excellent agreement with the behaviour
expected,~on~the general grounds detailed above, from quantum
mechanics: the decay constant $f_B$ predicted by~QCD sum rules is
strongly correlated with the value of the heavy-quark mass $m_b$
used as input. If all input parameters of the correlator except
for $m_b$---renormalization scales, $\alpha_{\rm s},$ quark
condensate, etc.---are kept fixed, we obtain a linear dependence
of $f_B$ on $m_b$ with negative slope, that is, an
anticorrelation:$$f_B(m_b)=\left(192.0-37\,\frac{m_b-4.247\;\mbox{GeV}}
{0.1\;\mbox{GeV}}\pm3_{\rm(syst)}\right)\mbox{MeV}\ .$$The above
strong (anti-) correlation between $f_B$ and $m_b$ enables us to
deduce an accurate value of $m_b$ from $f_B$ as starting point.
Feeding our average $f_B^{\rm LQCD}=(191.5\pm7.3)\;\mbox{MeV}$ of
recent findings for $f_B$ by some lattice-QCD collaborations
\cite{ETMC1,ETMC2,ALPHA,HPQCD1,HPQCD2,MILC} into our QCD sum-rule
investigation adopting the heavy--light correlator at
$O(\alpha_{\rm s}^2)$ accuracy yields the precise estimate
$m_b=(4.247\pm0.034)\;\mbox{GeV}.$

\section{Heavy--Light Two-Point Correlation Function and
(Borelized) QCD Sum Rule}This sum-rule study of the heavy
pseudoscalar $B_{(s)}$ mesons starts from the correlator
\cite{svz,aliev}~of two pseudoscalar currents
$j_5(x)\equiv(m_b+m)\,\bar q(x)\,{\rm i}\,\gamma_5\,b(x)$ of a $b$
quark and a light quark $q$ of
mass~$m$:$$\Pi\!\left(p^2\right)\equiv{\rm i}\int{\rm
d}^4x\exp({\rm i}\,p\,x)\left\langle0\left|
\mbox{T}\!\left(j_5(x)\,j^\dag_5(0)\right)\right|0\right\rangle.$$
Upon Borel transformation $\Pi\!\left(p^2\right)\to\Pi(\tau)$ to a
new ``Borel'' variable $\tau,$ the QCD sum rule sought~is obtained
by equating the results of evaluating this correlator at {\em QCD
level\/}, with the help of Wilson's operator product expansion
(OPE), and at {\em hadronic level\/}, by insertion of intermediate
hadron~states:\begin{align*}\Pi(\tau)&=f_B^2\,M_B^4
\exp\!\left(-M_B^2\,\tau\right)+\int\limits_{s_{\rm phys}}^\infty
\hspace{-.5ex}{\rm d}s \exp(-s\,\tau)\,\rho_{\rm hadr}(s)\\&=
\int\limits_{(m_b+m)^2}^\infty \hspace{-2.1ex}{\rm d}s
\exp(-s\,\tau)\,\rho_{\rm pert}(s,\mu)+\Pi_{\rm power}(\tau,\mu)\
,\end{align*}with the $B_{(s)}$ meson's mass $M_B$ and decay
constant $f_B$ defined by $(m_b+m)\,\langle0|\bar q\,{\rm
i}\,\gamma_5\,b|B\rangle=f_B\,M_B^2;$ the physical continuum
threshold, $s_{\rm phys}=(M_{B^*}+M_P)^2,$ is fixed by the beauty
vector meson's mass~$M_{B^*}$ and the mass $M_P$ of the lightest
pseudoscalar meson with appropriate quantum numbers, i.e., $\pi$
or~$K.$

For large $\tau,$ the contributions of the excited states to
$\Pi(\tau)$ decrease faster than the ground-state contribution, so
$\Pi(\tau)$ becomes saturated by the lowest state: the
large-$\tau$ behaviour of $\Pi(\tau)$ provides direct access to
ground-state features. However, analytic results for $\Pi(\tau)$
are found from a truncated OPE approximating $\Pi(\tau)$ well only
for $\tau$ not too large, where excited states still contribute
sizeably.

Excited-state contributions may be banished from $\Pi(\tau)$ by
assuming {\em quark--hadron duality\/}: all excited states'
contributions are counterbalanced by the perturbative contribution
above an {\em effective continuum threshold} $s_{\rm eff}(\tau),$
not to be confused with the physical continuum threshold: the
constant physical continuum threshold, $s_{\rm phys},$ is
determined by the masses of the lightest hadrons that may be
produced from the vacuum by the interpolating current whereas the
effective continuum threshold is a quantity intrinsic to the
sum-rule technique, with a lot of interesting and nontrivial
properties~\cite{lms_1}. Specifically, we have unambiguously shown
that the true effective threshold, defined by requiring it to
reproduce correctly the ground-state parameters, will exhibit a
dependence on the variable $\tau$ \cite{lms_new}. Applying duality
results in a relation, a {\em QCD sum rule\/}, between
ground-state observables~and~OPE:\begin{equation}\label{sr}
f_B^2\,M_B^4\exp\!\left(-M_B^2\,\tau\right)=
\int\limits_{(m_b+m)^2}^{s_{\rm eff}(\tau)}\hspace{-2.1ex}{\rm
d}s\exp(-s\,\tau) \,\rho_{\rm pert}(s,\mu)+\Pi_{\rm
power}(\tau,\mu)\ .\end{equation}Clearly, any evaluation of this
sum rule does not only require the knowledge of both
spectral~density $\rho_{\rm pert}(s,\mu)$ and nonperturbative
power corrections $\Pi_{\rm power}(\tau,\mu)$: in addition, we
have to formulate~or develop a criterion for determining $s_{\rm
eff}(\tau).$ Furthermore, we have to make sure that the OPE
exhibits a reasonable convergence; to this end, following
Ref.~\cite{jamin} we reorganize the perturbative expansion of
$\rho_{\rm pert}(s,\mu),$ derived in Ref.~\cite{chet} in terms of
the heavy quark's pole mass, in terms of the associated
$\overline{\rm MS}$ mass. The explicit results for $\rho_{\rm
pert}$ at three-loop level and $\Pi_{\rm power}$ may be found in
Refs.~\cite{chet,jamin}.

\section{Anticorrelation as Serendipity: Extracting the
$\overline{\rm MS}$ Mass $m_b$ of the Bottom Quark}
\begin{figure}[h]\begin{tabular}{cc}
\includegraphics[scale=.3782]{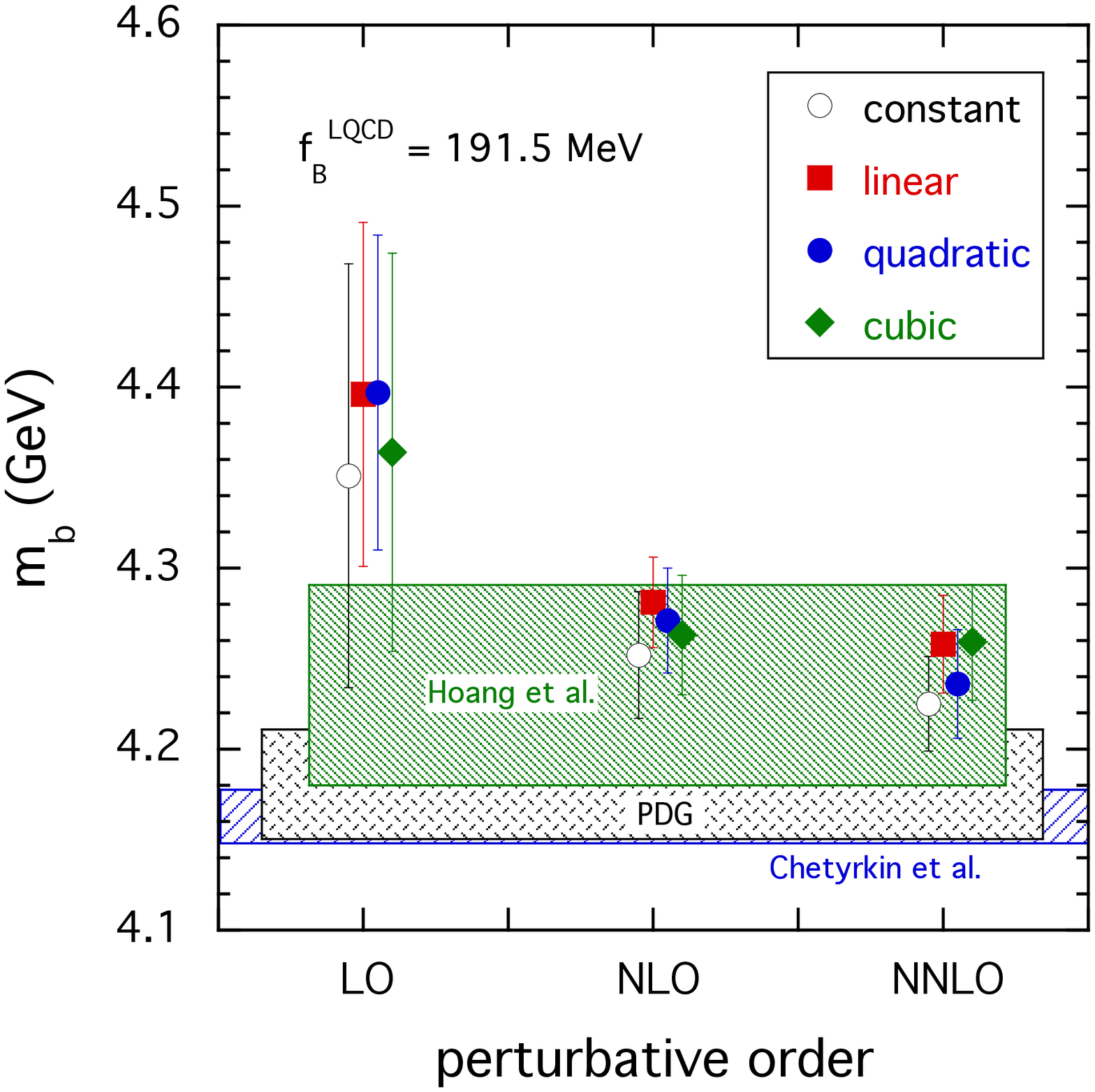}\hspace{1ex}&\hspace{1ex}
\includegraphics[scale=.3782]{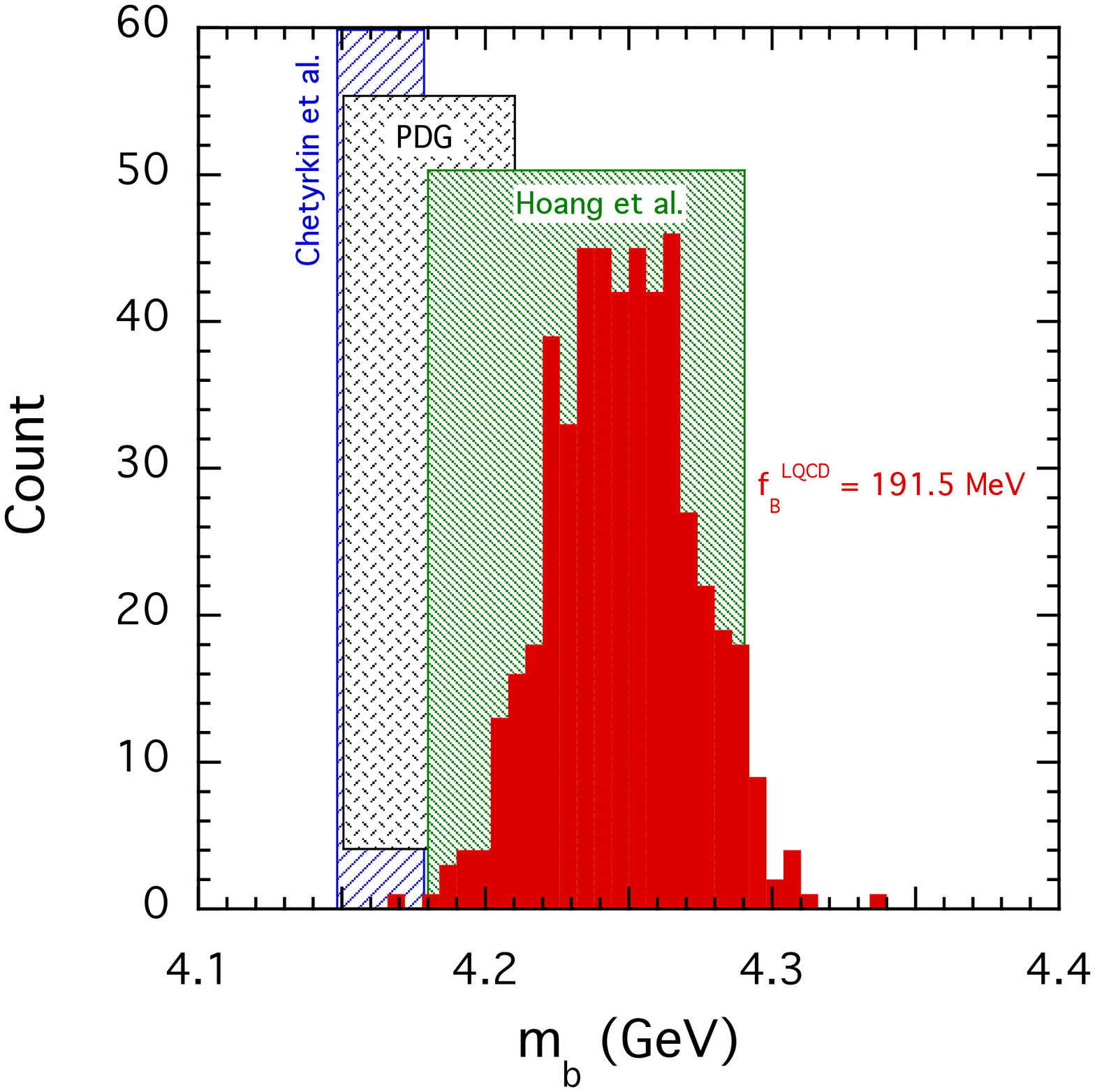}\\[1ex](a)&(b)\end{tabular}
\caption{Our findings for the $b$-quark mass
$m_b\equiv\overline{m}_b(\overline{m}_b),$ inferred from the
heavy--light QCD sum rule~(\protect\ref{sr}) by a bootstrap
analysis of all OPE errors for central value
$f_B=191.5\;\mbox{MeV}$ of the $B$-meson decay constant $f_B.$ (a)
Dependence of $m_b$ on the order of the perturbative expansion of
the correlator, indicated by ``LO,'' ``NLO,'' and ``NNLO,''
respectively. For comparison, the ($\pm1\,\sigma$) ranges of the
results found by the Particle Data Group (PDG) \cite{pdg}, by
Chetyrkin {\em et al.} \cite{mb1}, and by Hoang {\em et al.}
\cite{hoang}, for example, are represented by the shaded~areas.
(b) Distribution of $m_b$ from bootstrapping, adopting Gaussian
distributions for the OPE parameters except for the
renormalization scales $\mu$ and $\nu$ and uniform distributions
in the range $3\;\mbox{GeV}<\mu,\nu<6\;\mbox{GeV}$
for~$\mu$~and~$\nu.$}\label{Plot:mb}\end{figure}

The strong
sensitivity of $f_B$ and $f_{B_s}$ on the precise value of $m_b$
resulting from the QCD sum-rule approach allows us to invert our
line of thought and to derive an accurate prediction of
$m_b\equiv\overline{m}_b(\overline{m}_b)$ from (reasonably
accurate) lattice-QCD outcomes for $f_B$ and $f_{B_s}.$
Figure~\ref{Plot:mb} summarizes our findings, obtained from the
QCD sum rule (\ref{sr}) by applying our algorithms for fixing the
effective continuum threshold $s_{\rm eff}(\tau),$ which adopt a
polynomial {\em Ansatz\/} for $s_{\rm eff}(\tau)$ up to third
order (i.e., constant, linear, quadratic, or cubic dependence on
$\tau$). Figure \ref{Plot:mb}(a) depicts the resulting $m_b$
values for different orders taken into account in the perturbative
expansion of the correlator: Increasing its accuracy from $O(1)$
leading order (LO) to $O(\alpha_{\rm s})$ next-to-leading order
(NLO) diminishes central value and OPE error of $m_b$ from
$m_b^{\rm
LO}=(4.38\pm0.1_{\rm(OPE)}\pm0.020_{\rm(syst)})\;\mbox{GeV}$ to
$m_b^{\rm
NLO}=(4.27\pm0.04_{\rm(OPE)}\pm0.015_{\rm(syst)})\;\mbox{GeV}.$
Considering also the $O(\alpha_{\rm s}^2)$ next-to-next-to-leading
order (NNLO) has very little numerical impact: $m_b^{\rm
NNLO}=(4.247\pm0.027_{\rm(OPE)}\pm0.011_{\rm(syst)})\;\mbox{GeV}.$
Anyway, the extracted values of $m_b$ nicely show a kind of
convergence for increasing perturbative accuracy. The OPE error is
estimated by varying all OPE parameters in their ``usual''
intervals and both renormalization scales $\mu,\nu$ independently
in the range $3\;\mbox{GeV}<\mu,\nu<6\;\mbox{GeV}.$ For our final
result $m_b^{\rm NNLO},$ these quantities contribute
$14\;\mbox{MeV}$~($\mu,\nu$), $20\;\mbox{MeV}$ (quark condensate),
$7\;\mbox{MeV}$ (gluon condensate), $8\;\mbox{MeV}$ ($\alpha_{\rm
s}$) and $4\;\mbox{MeV}$ (light-quark mass), respectively, to the
total OPE error of $27\;\mbox{MeV},$ obtained by adding all the
individual contributions~in quadrature. The spread of $m_b$ values
for different $s_{\rm eff}(\tau)$ {\em Ans\"atze\/} is regarded as
systematic error~\cite{lms_fD} and amounts to $11\;\mbox{MeV};$
the lattice input $f_B=(191.5\pm7.3)\;\mbox{MeV}$ adds a Gaussian
error of $18\;\mbox{MeV}.$

\section{Summary of Main Results and Conclusions}This application
of QCD sum rules to the beauty-meson system provides some pivotal
insights:\begin{enumerate}\item Accepting the dependence of the
effective continuum threshold introduced when applying the notion
of quark--hadron duality on variables entering when performing
Borel transformations significantly improves the determination of
hadronic properties, by increasing the accuracy~of the duality
approximation and probing the intrinsic uncertainty of the QCD
sum-rule method.\item For beauty mesons, the sum-rule prediction
for $f_B$ is strongly correlated to the exact $m_b$ value:
$$\frac{\delta f_B}{f_B}\approx-8\,\frac{\delta m_b}{m_b}\ .$$
Realizing this behaviour, we use precise lattice-QCD results for
$f_{B_{(s)}}$ to extract the value of $m_b$ by combining the most
recent lattice-QCD findings for $f_B$ and $f_{B_s}$ with our
sum-rule~analysis:\begin{eqnarray}\label{ourmb}m_b=(4.247\pm0.027
_{\rm(OPE)}\pm0.018_{\rm(exp)}\pm0.011_{\rm(syst)})\;\mbox{GeV}\
.\end{eqnarray}Here, the OPE error arises from the errors of all
the OPE input parameters, the ``exp'' error~is a consequence of
the error in the QCD-lattice determination of $f_{B_{(s)}},$ and
the systematic error of the QCD sum-rule method inferred from the
spread of results when varying the {\em Ansatz\/} for the
effective continuum threshold is under control. Finally, adding
the errors in quadrature~yields$$m_b=(4.247\pm0.034)\;\mbox{GeV}\
.$$This implies, by the sum rule (\ref{sr}) from heavy--light
correlators evaluated at $O(\alpha_{\rm s}^2)$ accuracy,$$f_B=
\left(192.0\pm14.3_{\rm(OPE)}\pm3.0_{\rm(syst)}\right)\mbox{MeV}\
,\qquad f_{B_s}=\left(228.0\pm19.4_{\rm(OPE)}\pm4_{\rm(syst)}
\right)\mbox{MeV}\ .$$In view of the fact that the predicted value
of $m_b$ changes only marginally when increasing the correlator's
perturbative accuracy from $O(\alpha_{\rm s})$ to $O(\alpha_{\rm
s}^2),$ we do not expect that an inclusion of the at present
unknown $O(\alpha_{\rm s}^3)$ corrections will modify the
extracted value of $m_b$ substantially.\item Comparing our
prediction in Eq.~(\ref{ourmb}) with the other findings for $m_b$
available in the literature, we note agreement with
$m_b=(4.209\pm0.050)\;\mbox{GeV}$ from moment sum rules for
heavy--heavy correlators also at $O(\alpha_{\rm s}^2)$ accuracy
\cite{mb4}, with $m_b=(4.235\pm0.055_{\rm(pert)}\pm0.003
_{\rm(exp)})\;\mbox{GeV}$ from a renormalization-group-improved
next-to-next-to-leading-logarithmic-order discussion of $\Upsilon$
sum rules \cite{hoang} as well as, within two standard deviations,
with the Particle Data Group average
$m_b=(4.18\pm0.03)\;\mbox{GeV}$ \cite{pdg} but an evident
disagreement with $m_b=(4.163\pm0.016)\;\mbox{GeV}$ \cite{mb1} and
$m_b=(4.171\pm0.009)\;\mbox{GeV}$ \cite{dominguez} due to sum
rules using heavy--heavy correlators at $O(\alpha_{\rm s}^3)$
accuracy; we doubt that $O(\alpha_{\rm s}^3)$ corrections to the
heavy--light sum rule can restore~agreement.\end{enumerate}In
conclusion, let us emphasize that {\em properly formulated\/}
Borel QCD sum rules for heavy--light correlators form competitive
tools both for reliable determinations of heavy-meson observables
and for the extraction of basic QCD parameters by exploiting the
results of lattice QCD and experiment.

\vspace{4ex}\noindent{\bf Acknowledgments.} D.M.\ was supported by
the Austrian Science Fund (FWF), project no.~P22843.


\begin{thebibliography}{99}
\bibitem{ETMC1}P.~Dimopoulos {\em et al.} (ETM Collaboration), JHEP
{\bf 1201} (2012) 046.
\bibitem{ETMC2}N.~Carrasco {\em et al.} (ETM Collaboration),
PoS({\bf Lattice 2012})104 (2012); PoS({\bf ICHEP2012})428 (2012).
\bibitem{mb_Gimenez}V.~Gimenez, L.~Giusti, G.~Martinelli, and
F.~Rapuano, JHEP {\bf 03} (2000) 018.
\bibitem{mb_UKQCD}C.~McNeile, C.~Michael, and G.~Thompson (UKQCD
Collaboration), Phys.~Lett.~B {\bf 600} (2004) 77.
\bibitem{ALPHA}F.~Bernardoni {\em et al.} (ALPHA Collaboration),
Nucl.~Phys.~Proc.~Suppl.~{\bf 234} (2013) 181.
\bibitem{mb4}J.~H.~K\"uhn and M.~Steinhauser, Nucl.~Phys.~B {\bf
619} (2001) 588.
\bibitem{mb1}K.~G.~Chetyrkin {\em et al.}, Phys.~Rev.~D {\bf 80}
(2009) 074010.
\bibitem{mb2}C.~McNeile {\em et al.} (HPQCD Collaboration),
Phys.~Rev.~D {\bf 82} (2010) 034512.
\bibitem{hoang}A.~Hoang, P.~Ruiz-Femenia, and M.~Stahlhofen, JHEP
{\bf 1210} (2012) 188.
\bibitem{lms2013}W.~Lucha, D.~Melikhov, and S.~Simula, Phys.~Rev.~D
{\bf 88} (2013) 056011.
\bibitem{lms_qcdvsqm}W.~Lucha, D.~Melikhov, and S.~Simula,
Phys.~Lett.~B {\bf 687} (2010) 48; Phys.~Atom.~Nucl.~{\bf 73}
(2010) 1770.
\bibitem{svz}M.~A.~Shifman, A.~I.~Vainshtein, and V.~I.~Zakharov,
Nucl.~Phys.~B {\bf 147} (1979) 385.
\bibitem{aliev}T.~M.~Aliev and V.~L.~Eletsky, Yad.~Fiz.~{\bf 38}
(1983) 1537.
\bibitem{nar2001}S.~Narison, Phys.~Lett.~B {\bf 520} (2001) 115.
\bibitem{jamin}M.~Jamin and B.~O.~Lange, Phys.~Rev.~D {\bf 65}
(2002) 056005.
\bibitem{lms_fB}W.~Lucha, D.~Melikhov, and S.~Simula, J.~Phys.~G
{\bf 38} (2011) 105002.
\bibitem{nar2012}S.~Narison, Phys.~Lett.~B {\bf 718} (2013) 1321.
\bibitem{chet}K.~G.~Chetyrkin and M.~Steinhauser,
Phys.~Lett.~B {\bf 502} (2001) 104; Eur.~Phys.~J.~C {\bf 21}
(2001) 319.
\bibitem{HPQCD1}H.~Na {\em et al.} (HPQCD Collaboration),
Phys.~Rev.~D {\bf 86} (2012) 034506.
\bibitem{HPQCD2}C.~McNeile {\em et al.} (HPQCD Collaboration),
Phys.~Rev.~D {\bf 85} (2012) 031503(R).
\bibitem{MILC}A.~Bazavov {\em et al.} (Fermilab Lattice and MILC
Collaborations), Phys.~Rev.~D {\bf 85} (2012) 114506.
\bibitem{lms_1}W.~Lucha, D.~Melikhov, and S.~Simula, Phys.~Rev.~D
{\bf 76} (2007) 036002; Phys.~Lett.~B {\bf 657} (2007) 148;
Phys.~Atom.~Nucl.~{\bf 71} (2008) 1461; Phys.~Lett.~B {\bf 671}
(2009) 445; D.~Melikhov, Phys.~Lett.~B {\bf 671} (2009) 450.
\bibitem{lms_new}W.~Lucha, D.~Melikhov, and S.~Simula, Phys.~Rev.~D
{\bf 79} (2009) 096011; J.~Phys.~G {\bf 37} (2010) 035003;
W.~Lucha, D.~Melikhov, H.~Sazdjian, and S.~Simula, Phys.~Rev.~D
{\bf 80} (2009) 114028.
\bibitem{pdg}J. Beringer {\em et al.} (Particle Data Group),
Phys.~Rev.~D {\bf 86} (2012) 010001.
\bibitem{lms_fD}W.~Lucha, D.~Melikhov, and S.~Simula, Phys.~Lett.~B
{\bf 701} (2011) 82.
\bibitem{dominguez}S.~Bodenstein {\em et al.}, Phys.~Rev.~D {\bf
85} (2012) 034003.\end{thebibliography}
\end{document}